\documentclass[5p,twocolumn,times,number]{elsarticle}

\usepackage{graphicx}
\usepackage{amsmath}   

\begin{document}

\begin{frontmatter}

\title{The OPERA global readout and GPS distribution system}

\author[add1]{J.~Marteau\corref{cor}}
\ead{j.marteau@ipnl.in2p3.fr}
\author[add2]{the OPERA collaboration}

\cortext[cor]{Corresponding author}

\address[add1]{IPNL, Universit\'e de Lyon, Universit\'e Lyon 1, CNRS/IN2P3, 4 rue E. Fermi 69622 Villeurbanne cedex, France}

\begin{abstract}
  OPERA is an experiment dedicated to the observation of $\nu_\mu$ into $\nu_\tau$ oscillations in appearance mode using a pure  $\nu_\mu$ beam (CNGS) produced at CERN and detected at Gran Sasso.
  The experiment exploits a hybrid technology with emulsions and
electronics detectors~\cite{opera}. The OPERA readout is performed through a triggerless, continuously running, distributed and highly available system. Its global architecture is based on Ethernet-capable smart sensors with microprocessing and network interface directly at the front-end stage. An unique interface board is used for the full
detector reading out ADC-, TDC- or Controller-boards. 
  All the readout channels are synchronized through a GPS-locked common bidirectional clock distribution system developped on purpose in a PCI format. It offers a second line to address all channels and the off-line synchronization with the CNGS to select the events. 
\end{abstract}


\end{frontmatter}

\section{Global data readout architecture}

OPERA is optimized for the $\tau$ neutrino detection. Detector's target consists of lead-emulsions bricks walls followed by scintillator tracker planes (Target Tracker referred to as TT) to trigger and locate the events and it is complemented by two muon magnetic spectrometers (drift tubes referred to as HPT and RPC tracker). The overall data rate is dominated by cosmics and ambient radioactive background while the neutrino interactions are well localized in time in correlation with the CNGS beam spill. The synchronization with the CNGS beam spill is done off-line via the GPS. The detector remains sensitive during the inter-spill time and the DAQ system runs continuously in a triggerless mode.\\
The data readout system has been designed to sort the data through Ethernet at the earliest stage of each sub-detector. The global DAQ is build like a standard Ethernet network whose nodes are the ``Ethernet Controller Mezzanines'' (ECM) plugged on Controller Boards (CB). These boards are designed to interface and control each sub-detector specific F/E electronics, to sort the data to the Event Building WorkStation (EBWS), to handle monitoring and slow control from the Global Manager through the same Ethernet processors.\\ \textit{Target Tracker} The readout of the TT PMTs (Hamamatsu 64 pixels H8804-mod1) is performed via multi-channels front-end chips designed at LAL the signals of which are digitized and pre-processed on a custom motherboard developed at the IPNL. The F/E chips are self-triggering and allow to compensate the pixel-to-pixel gain spread of the MaPMT. A common threshold is set on a chip, which makes the OR of the 32 channels to provide a trigger output signal whenever one channel at least is over threshold.\\ 
\textit{Spectrometer} Each spectrometer has two sensitive elements: 22 RPC planes and 3 pairs of drift tubes stations located upstream, in the middle and downstream of a dipolar magnet. The electronics of these 2 trackers is placed on the top of the experiment. It consists in VME-like boards collected in VME crates and power supply stations. The F/E electronics of one RPC plane is divided into 9 F/E boards readout serially by one Controller Board that receives the trigger signals (defined by the fast OR of 32 channels). Each drift tube station is readout by a TDC board which receives a common stop from the OR of two nearby RPC planes.\\ 
\textit{Computing} The EBWS consists of commercial PC's receiving data from the Ethernet network. The distributed client/server software is based on the CORBA (Common Object Request Broker Architecture) standard which is a well established Object Oriented application~\cite{corba}. It offers a global framework which links the applications running on the sensors, on the EBWS and on the Manager. The on-line data storage is performed by a database cluster (Oracle 11.1 RAC) with two dedicated servers.\\
\textit{Network} The global DAQ network is divided into 2 parts~: the Ethernet network from the EBWS to each Ethernet Controller (used for data transfer but also for detector configuration, monitoring \& slow control) and the clock distribution system from the Central Clock Unit synchronized on the GPS to each sensor (used for the synchronization of all the nodes local clocks)

\section{Ethernet Controller Mezzanine}
\label{sec:ecm}

\begin{figure}
\centering
\includegraphics[width=7.5cm]{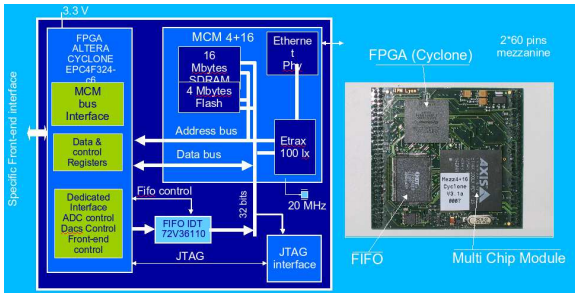}
\caption{Ethernet Controller Module: schematics and physical view.}\label{fig:camerop}
\end{figure}

The aim of the ECM is to provide a {\bf common interface} between the F/E of the various sub-detectors and the overall DAQ system. The physical form of the mezzanine has been chosen for the ease of integration (see Figure~\ref{fig:camerop}). The mezzanine can be seen as an external component to be plugged on the Controller Boards and that can be (re-)configured to the specifications of each sub-detectors~\cite{claude,girerd,opera_note_1}.
The Ethernet controller is the core of the board. It includes a sequencer (FPGA from the Cyclone ALTERA family), an external FIFO and a microprocessor (ETRAX 100LX from AXIS) with an Ethernet interface.\\ 
The FPGA performs the sequencing of the readout (clocks, R/O registers, digital I/O), the local data pre-processing (zero suppression, event validation through external trigger on request), event time stamping (via a local fine counter at 100 MHz synchronized with the global distributed clock), data transfer to the external FIFO, interface with the Ethernet processor. 
The Ethernet processor is an ASIC based on the ETRAX 100LX from AXIS. It is a 32-bit RISC CPU with Linux 2.4 operating system and Ethernet interface supporting data transfer rates up to 200 Mbits/s~\cite{axis}. The chip is embedded in a Multi Chip Module (MCM) which includes in a single chip the ETRAX100lx core plus 4 Mbytes of flash memory, 16 Mbytes of SDRAM and an Ethernet transceiver. The MCM gets the data in a time scheduled job from the FIFO, performs local processing and transfers the data to the EBWS.

\section{Clock distribution system}
\label{sec:clock}

\begin{figure}
\centering
\includegraphics[width=0.9\linewidth,height=4cm]{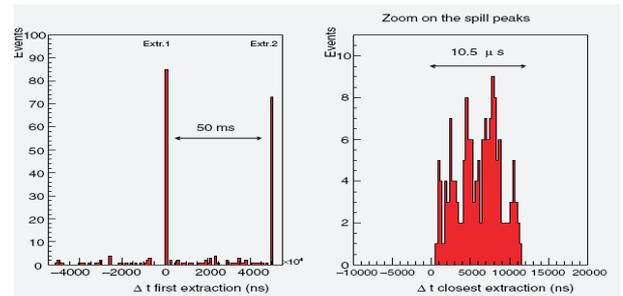}\
\caption{CNGS beam time structure seen in OPERA}
\label{fig:clock_2} 
\end{figure}

A global clock is mandatory to synchronize all the nodes of the distributed system. The general features of the system are close to Ref.\cite{antares_clk}, in particular the bi-directionnality of the system allows the control of the signal reception and the measurement of the propagation time with acknowledgement signals.
The clock distribution system starts from the GPS control unit which synchronizes a 20MHz clock with the signal of the GPS and sends the clock + encoded commands via an optical fibre. The signal is then converted into electrical format and distributed to the "clock master cards" through M-LVDS bus. Each of these cards deserialize the commands and the clock, and distribute both of them to the clock unit of each controller board through another M-LVDS bus.
The PPS sent to each sensor is used to reset the local fine counters after correction of the time propagation delay. Since the PPS transmission time is recorded and locked with the GPS, the absolute event timestamp is assigned as the sum of that time and the value of the local 100MHz clock running on the ECM. This timestamp accuracy allows monitoring the beam spill structure (Figure~\ref{fig:clock_2}) and to perform physics studies in parallel to the neutrino oscillations using particles ToF (cosmic rays physics, atmospheric neutrinos etc).
A crucial point to extract the beam events is the intercalibration between CERN and LNGS GPS systems. Both have comparable performance (<100 ns accuracy) and their single units are in both cases based on a GPS system + Rb clock although build by different manufacturers (Symmetricom for CERN, ESAT for LNGS). Dedicated tests were performed to check for relative offsets and time stability of the two systems~\cite{Serrano:2006zz}. The two systems are able to track each other within $\pm 23$ns over long periods. 

\section{Conclusion}
The OPERA global readout system is the first fully Ethernet-based data acquisition system using the ``smart sensors'' concept. Commissioned during the first CNGS run in summer 2006, it is running continuously with a high availability and efficiency. During the 2008 physics run, around 10,000 neutrino interactions have been recorded in time coincidence with the CNGS, among which 1,700 interactions in the target. Emulsions analysis is being performed by automatic scanning stations~\cite{ieva}.


\end{document}